\documentclass[aps,prc,superscriptaddress,showpacs]{revtex4}
\usepackage{bm}

\newcommand{\vct}[1]{{\bf #1}}
\newcommand{\svct}[1]{\vct{#1}}

\newcommand{\ket}[1]{|#1\rangle}
\newcommand{\mtrix}[3]{\langle \,#1\,|#2|\,#3\,\rangle}
\newcommand{\fslash}[1]{\!\not\!#1}
\newcommand{\kf}{k_{\text{F}}}
\newcommand{\ekf}{E_{\text{F}}}
\newcommand{\kfp}{k_{\text{p}}}
\newcommand{\kfn}{k_{\text{n}}}
\newcommand{\vf}{v_{\text{F}}}
\newcommand{\mitg}{{\mit\Gamma}}
\newcommand{\thp}[1]{\theta^{\text{(p)}}_{\vct{#1}}}
\newcommand{\thn}[1]{\theta^{\text{(n)}}_{\vct{#1}}}
\newcommand{\ep}[1]{E_{\vct{#1}}}
\newcommand{\rmtrix}[3]{\langle \,#1\,||\,#2\,||\,#3\,\rangle}
\newcommand{\kfi}{k_i}
\newcommand{\thi}[1]{\theta^{\text{(i)}}_{\svct{#1}}}
\newcommand{\nnbar}{\text{N}\overline{\text{N}}}

\begin{document}

\title{The Gamow-Teller States in Relativistic Nuclear Models}

\author{Haruki Kurasawa}
\affiliation{Department of Physics, Faculty of Science, Chiba University,
Chiba 263-8522, Japan}

\author{Toshio Suzuki}
\affiliation{Department of Applied Physics, Fukui University,
Fukui 910-8507, Japan\\
RIKEN, 2-1 Hirosawa, Wako-shi, Saitama 351-0198, Japan
}

\author{Nguyen Van Giai}
\affiliation{%
Institut de Physique Nucl$\acute{e}$aire,
CNRS-IN2P3, 91406 Orsay Cedex, France
}

\begin{abstract}
The Gamow-Teller(GT) states are investigated in
relativistic models.
The Landau-Migdal(LM) parameter is
introduced in the Lagrangian as a contact term with the pseudo-vector
coupling.
In the relativistic model the total GT strength in the nucleon space
is quenched by about 12\% in nuclear matter and by about 6\% in
finite nuclei, compared with the one of the Ikeda-Fujii-Fujita sum
rule. The quenched amount is taken by nucleon-antinucleon excitations
in the time-like region.
Because of the quenching, the relativistic model
requires a larger value of the LM parameter 
than non-relativistic models in describing
the excitation energy of the GT state. 
The Pauli blocking terms are not important for the description of
the GT states.
\end{abstract}


\maketitle

\section{Introduction}

For the last 30 years it has been shown
that relativistic models work very well
phenomenologically to explain various nuclear phenomena\cite{swring}.
Most of them assume that the nucleus is a relativistic system composed
of the Dirac particles in the Lorentz scalar and vector potentials.

In the present paper,
we study the excitation energy and strength of
the Gamow-Teller(GT) states in the relativistic models.
As far as the authors know, the GT states have not been studied
in detail so far\cite{conti}.
We will discuss those mainly in nuclear matter, since we can obtain 
analytic expressions of the excitation energy and strength
which make clear the structure of the relativistic
model and the difference between the
relativistic and non-relativistic models. 

In the next section we will present our relativistic framework to
discuss the GT states. The Landau-Migdal(LM) parameter will be
introduced in the Lagrangian as a contact term to take
into account particle-hole correlations.
In the section \ref{sec_tr} the transverse correlation function will be
calculated explicitly, from which  an analytic expression of the
excitation energy will be obtained in section \ref{sec_ex}.
In section \ref{sec_gt}, the GT strength 
will be calculated. We will show
that the total GT strength is quenched by about
12\% in nuclear matter and by about 6\% in finite nuclei, compared with
the non-relativistic sum rule value. The quenched strength is taken by
the nucleon-antinucleon excitations in the time-like region,
which can not be excited with
charge-exchange reactions. 
Effects of the Pauli blocking terms on the
excitation energy and strength will be shown to be negligible in
section \ref{sec_ex} and \ref{sec_gt}.
In section \ref{sec_pi}, we will show
that the way to add the LM parameter to the relativistic meson
propagator, which was frequently used in a description of high-momentum
transfer reactions\cite{h}, can not describe the GT states.
The last section will be devoted to a brief summary of the present work.

\section{ Relativistic Model} \label{sec_rel}

We assume that the mean field is provided by the Lorentz scalar and
vector potential.
The RPA correlations are described using
the basis given in this mean field, and are assumed to be induced
through the Lagrangian:
\begin{eqnarray}
 {\cal L}=
 -\,g_\pi
 \overline{\psi}\mitg^\mu_i\psi\,\partial_\mu\pi_i
 +\frac{g_5}{2}\,
 \overline{\psi}\mitg^\mu_i\psi\,
 \overline{\psi}\mitg_{\mu i}\psi \label{L}
\end{eqnarray}
with
\[
\mitg^\mu_i=\gamma_5\gamma^\mu\tau_i \,, \ \ \ 
g_\pi=\frac{f_\pi}{m_\pi} , \ \ \
g_5=\left(\frac{f_\pi}{m_\pi}\right)^2g'.
\]
The first term stands for the usual PV coupling between the pion and
nucleon, and the second term is to take into account the LM parameter
$g'$\cite{mosel}.
Although
it is model-dependent how to introduce $g'$ in the relativistic
model, we will show  that the above Lagrangian yields the known
expression for the excitation energy of the GT state in the
non-relativistic limit.
The first term, in fact, is not relevant for
the GT states in nuclear matter. It is, however, kept
in order to show later that if the LM parameter is put
into the meson propagator, we can not describe the GT state.

For the Lagrangian Eq.(\ref{L}), the RPA correlation
function ${\mit\Pi}_{\rm RPA}$ is written in terms of the mean field
one ${\mit\Pi}$\cite{ks1}, 
\begin{eqnarray}
{\mit\Pi}_{\rm RPA}(\mitg_A,\mitg_B\,)
&=&{\mit\Pi}(\mitg_A,\mitg_B\,)
+\chi_\pi(q)\,
{\mit\Pi}(\mitg_A,\mitg_i\!\cdot\!q\,)\,
{\mit\Pi}_{\rm RPA}(\mitg_{i}\!\cdot\!q,\mitg_B\,)
+\chi_5\,
{\mit\Pi}(\mitg_A,\mitg^\mu_i\,)\,
{\mit\Pi}_{\rm RPA}(\mitg_{\mu i},\mitg_B\,),\label{rpa}
\end{eqnarray}
where the following notations are employed,
\[
\chi_\pi(q)=
\frac{g_\pi^2}{(2\pi)^3}\frac{1}{m_\pi^2-q^2-i\varepsilon}\,,\quad
\chi_5=
\frac{g_5}{(2\pi)^3}.
\]
For isospin-dependent excitations, 
the mean field correlation function is given by
\begin{eqnarray}
{\mit\Pi}( \mitg_\alpha , \mitg_\beta \,) 
&=&
-\, \frac{1}{2\pi i}\int\!\!d^4p\,
{\rm Tr}_\sigma{\rm Tr}_\tau\Bigl( 
\mitg_\alpha G_{\rm F}(p+q)\mitg_\beta G_{\rm D}(p)
+\mitg_\alpha G_{\rm D}(p+q)\mitg_\beta G_{\rm F}(p)
\Bigr. \nonumber\\
& &
 \phantom{
-\, \frac{1}{2\pi i}\int\!\!d^4p\,{\rm Tr}_\sigma{\rm Tr}_\tau
}
+\Bigl.\mitg_\alpha G_{\rm D}(p+q)\mitg_\beta G_{\rm D}(p)
+\mitg_\alpha G_{\rm F}(p+q)\mitg_\beta G_{\rm F}(p)
\Bigr),\label{hpi}
\end{eqnarray}
where we have defined the isospin operator $\tau_\alpha$:
\[
\tau_\pm=
\frac{\tau_x \pm i\tau_y}{\sqrt{2}}\, \ \ ,\qquad
\tau_0=\tau_z,
\]
and the propagator:
\[
 G_{\rm H}(q)=G_{\rm F}(q)+G_{\rm D}(q)\,,\quad
 G_{\rm D}(q)=G(\kfp\,;\,q)\frac{1-\tau_z}{2}
 +G(\kfn\,;\,q)\frac{1+\tau_z}{2},
\]
with
\[
G_{\rm F}(q)= \frac{\fslash{q}+M^\ast}
{q^2-M^{\ast\,2}+i\varepsilon}\,,\qquad
G(k_i\,;\,q)=
\frac{i\pi}{E_{\svct{q}}}\left(\fslash{q}+M^\ast\right)
\delta(q_0-E_{\svct{q}} )\,\theta_{\vct{q}}^{(i)}, \ \ \ (i={\rm p, n}).
\]
Here, we have also used the abbreviation for the step function:
$\theta_{\vct{q}}^{(i)}=\theta(k_i -|\vct{q}|)$\,, 
$\kfp$ and $\kfn$ being the Fermi momentums of
the protons and neutrons, respectively.
Moreover, $E_{\svct{q}}$ is equal to $\sqrt{M^{\ast 2}+\vct{q}^2}$,
where the Lorentz scalar potential is included
in the nucleon effective mass $M^\ast$. The Lorentz vector potential 
does not show up explicitly in the present discussion of the nuclear
matter. In Eq.(\ref{hpi}), the first three terms are
density-dependent, including the
Pauli blocking terms, while the last one is density-independent and
divergent. The last term 
is usually neglected\cite{chin}, but we keep it
for later discussions.
Effects of the Pauli blocking terms on the excitation energy and
strength of the GT state will be also discussed later. 

For the $\tau_\pm$ excitations, the RPA correlation function in
Eq.(\ref{rpa})  is described as
\begin{eqnarray}
{\mit\Pi}_{\rm RPA}(\mitg^a_+\,, \mitg^b_- )
=[U^{-1}]^{ab'}
{\mit\Pi}({\mit\Gamma}_{b'+}\,, \mitg^b_- ),
\end{eqnarray}
where $U$ denotes the dimesic function of the $5\times 5$ matrix: 
\[
U^{ab}=
g^{ab}-\chi_b{\mit\Pi}({\mit\Gamma}^a_+\,, {\mit\Gamma}^b_-\,)
\]
with the notations for
$a=-1,\,0,\,\cdots,3$,
\[
{\mit\Gamma}^a_\pm =\gamma_5\gamma^a\tau_\pm \,,\qquad
\gamma^a
=\left\{
\begin{array}{l}
 \gamma\!\cdot\!q\\
\noalign{\vskip4pt} 
 \gamma^\mu\\
\end{array}
\right. 
,\qquad
\chi_a
=\left\{
\begin{array}{ll}
 \chi_\pi   &\,,\  a=-1\\
\noalign{\vskip8pt} 
 \chi_5 &\,,\  a=\mu. \\
\end{array}
\right. 
\]
In the above equation, $g^{ab}$ is defined as $g^{ab}=1(a=b=-1),\ 
=g^{\mu\nu}(a=\mu, b=\nu)$ and $g^{-1\mu} =g^{\mu -1}=0$.

The calculation of the mean field correlation function is
straightforward. Separating
${\mit\Pi}( {\mit\Gamma}^a_+ , {\mit\Gamma}^b_-\,)$ into the
density-dependent
and independent parts:
\begin{eqnarray}
{\mit\Pi}( {\mit\Gamma}^a_+ , {\mit\Gamma}^b_-\,)=
 {\mit\Pi}_{\rm D}( {\mit\Gamma}^a_+ , {\mit\Gamma}^b_-\,)
+ {\mit\Pi}_{\rm F}( {\mit\Gamma}^a_+ , {\mit\Gamma}^b_-\,),
\end{eqnarray}
they are obtained as,
\begin{eqnarray}
{\mit\Pi}_{\rm D}( {\mit\Gamma}^a_+ , {\mit\Gamma}^b_-\,)
&=&
\int\!\!d^4p\,\frac{\delta(p_0-E_{\svct{p}} )}{E_{\svct{p}}}
\left(
\frac{t^{ab}(p,q)}{(p+q)^2-M^{\ast2}+i\varepsilon}
\,\thn{p}
+\, \frac{t^{ab}(p,-q)}{(p-q)^2-M^{\ast2}+i\varepsilon}
\,\thp{p}
\right) \nonumber\\
 \noalign{\vskip4pt} 
& &
+\, i\pi
\int\!\!d^4p\,\frac{\delta(p_0-E_{\svct{p}} )\,
\delta(p_0+q_0-E_{\svct{p}+\svct{q}} )}
{E_{\svct{p}} E_{\svct{p}+\svct{q}} }
\,t^{ab}(p,q)\,\thn{p}\thp{p+q},
\label{mf}\\
 \noalign{\vskip4pt} 
{\mit\Pi}_{\rm F}( {\mit\Gamma}^a_+ , {\mit\Gamma}^b_-\,)
&=& \frac{1}{i\pi}\int\!\!d^4p\,
\frac{t^{ab}(p,q)}
{
\left(p^2-M^{\ast2}+i\varepsilon\right)
\left((p+q)^2-M^{\ast2}+i\varepsilon\right)
},\label{mff}
\end{eqnarray}
where $t^{ab}(p,q)$ is given by
\begin{eqnarray}
t^{\mu\nu}(p,q)&=&4\Bigl(
  g^{\mu\nu}\left(M^{\ast2}+p^2+p\!\cdot\!q \right)
-2p^\mu p^\nu -p^\mu q^\nu-p^\nu q^\mu  
\Bigr),\label{t1} \\
\noalign{\vskip4pt}
t^{-1\nu}(p,q)&=& q_\mu t^{\mu\nu}(p,q)=4\Bigl(
  q^{\nu}\left(M^{\ast2}+p^2 \right)
-p^\nu\left( q^2+2p\!\cdot\!q \right)
\Bigr),\label{t2}
\\
\noalign{\vskip4pt}
t^{-1-1}(p,q)&=&4\Bigl(
  q^2\left(M^{\ast2}+p^2 \right)
-p\!\cdot\!q\left( q^2+2p\!\cdot\!q \right)
\Bigr).\label{t3}
\end{eqnarray}

When we define the three dimensional axes as
$q^\mu=(q_0,\,q_x,\,0,\,0)$, we can show that
${\mit\Pi}( \mitg^a_- , \mitg^b_+\,)$ has a structure as shown in
Table \ref{table1}--(a), where open boxes mean
${\mit\Pi}( \mitg^a_- , \mitg^b_+\,)$ to be non-zero.
Thus, the transverse parts of ${\mit\Pi}( \mitg^a_- , \mitg^b_+\,)$
are decoupled from the pion($a=-1$)--, time($a=0$)-- and
longitudinal($a=1$) ones. In the present paper,
we are interested in the GT states excited
at $\vct{q}=0$. In this case, the longitudinal part is
also decoupled from the pion- and time-component(PT), as in
Table \ref{table1}--(b).
Consequently, the determinant of
the dimesic function is factorized into four parts,
\begin{eqnarray}
{\rm det}\ U =-\left(D_{\rm T}\right)^2D_{\rm L}D_{\rm PT},
\end{eqnarray}
where the transverse, longitudinal and PT dimesic functions are
written as, respectively,
\begin{eqnarray}
D_{\rm T}&=&1+\chi_5\,{\mit\Pi}(\mitg^{2}_+\,, \mitg^{2}_-\,), \qquad
 D_{\rm L}=D_{\rm T},
 \label{t} \\
\noalign{\vskip6pt}
D_{\rm PT}
&=&\Bigl(
    1-\chi_\pi {\mit\Pi}(\mitg^{-1}_+\,, \mitg^{-1}_-\,)
   \Bigr)
\Bigl(
    1-\chi_5 {\mit\Pi}(\mitg^{0}_+\,, \mitg^{0}_-\,)
   \Bigr)
-\chi_5\chi_\pi
{\mit\Pi}(\mitg^{-1}_+\,, \mitg^{0}_-\,)
{\mit\Pi}(\mitg^{0}_+\,, \mitg^{-1}_-\,).\label{l}
\end{eqnarray}

\section{The Transverse Correlation Function} \label{sec_tr}

In this section we will derive more explicit form of the 
transverse correlation function at $\vct{q}=0$
which is used for description of the GT states.
First we will calculate the real and imaginary part of the
density-dependent transverse correlation function separately, and next
the density-independent part.

According to Eq.(\ref{mf}),
the real part of the density-dependent transverse correlation function
is written as
\begin{eqnarray}
{\rm Re}\ {\mit\Pi}_{\rm D}(\mitg^{2}_+\,, \mitg^{2}_-\,)=
J^{22}(\kfn,q_0) +J^{22}(\kfp,-q_0),\label{retra}
\end{eqnarray}
where $J^{22}(k_i, q)$ represents
\begin{eqnarray}
J^{22}(k_i,q)=
\int\!\!d^4p\,\frac{\delta(p_0-E_{\svct{p}} )}{E_{\svct{p}}}
\frac{t^{22}(p,q)}{(p+q)^2-M^{\ast2}}
\,\theta_{\vct{p}}^{(i)}.
\end{eqnarray}
Since $t^{22}$ at $\vct{q}=0$ is given by
\begin{equation}
t^{22}(p,q)
=-\,4
\Bigl(2M^{\ast2}+E_{\svct{p}}q_0+2p_y^2
\Bigr),\label{t22}
\end{equation}
the real part Eq.(\ref{retra}) is described as
\begin{eqnarray}
{\rm Re}\,{\mit\Pi}_{\rm D}( \mitg^2_+ , \mitg^2_-\,)
&=& 
-\,\frac{4}{q_0}
\int\!d^3p\,
\frac{M^{\ast2}+p_y^2}{\ep{p}^2}\left(\thn{p} - \thp{p} \right)
\nonumber\\
& &
-4\int\!d^3p\,
\frac{\vct{p}^2-p_y^2}{\ep{p}^2}
\left(
\frac{\thn{p}}{2\ep{p}+q_0}
+\frac{\thp{p}}{2\ep{p}-q_0}
\right). \label{eq_re_pauli}
\end{eqnarray}
The imaginary part of the density-dependent correlation function
at $\vct{q}=0$ is given by Eq.(\ref{mf}) as
\begin{eqnarray*}
{\rm Im}\, {\mit\Pi}_{\rm D}( \mitg^a_+ , \mitg^b_-\,)
&=&
-\,\pi
\int\!\!d^4p\,\frac{\delta(p_0-\ep{p} )}{\ep{p}}
\Bigl[\,
 t^{ab}(p,q)\,\delta( q_0^2+2p_0q_0 )\,\thn{p}
 \Bigr. \\
& &
\phantom{
-\,\pi
} 
+\,t^{ab}(p,-q)\,\delta( q_0^2-2p_0q_0 )\,\thp{p}
-\frac{\delta(q_0)}{\ep{p} }
\,t^{ab}(p,q)\,\thn{p}\,\thp{p}
\,\Bigr].
\end{eqnarray*}
Using
\[
\delta( q_0^2\pm 2p_0q_0 )
=\frac{1}{2p_0}\Bigl( \delta(q_0)+\delta(q_0\pm 2p_0) \Bigr),
\]
the above equation is rewritten as
\begin{eqnarray*}
{\rm Im}\, {\mit\Pi}_{\rm D}( \mitg^a_+ , \mitg^b_-\,)
&=&
-\,\pi\,\delta(q_0)
\int\!\!d^4p\,\frac{\delta(p_0-\ep{p} )}{2\ep{p}^2}
\Bigl( t^{ab}(p,q)\,\thn{p}+t^{ab}(p,-q)\,\thp{p}
-\,2t^{ab}(p,q)\,\thn{p}\,\thp{p}
\Bigr)
+R_{\nnbar}.
\end{eqnarray*}
The last term $R_{\nnbar}$ comes from the  N--$\overline{\rm N}$
excitations,
\[
 R_{\nnbar}(q_0)
=-\,\pi
\int\!\!d^4p\,\frac{\delta(p_0-\ep{p} )}{2\ep{p}^2}
\Bigl(
 t^{ab}(p,q)\,\delta(q_0+2\ep{p} )\,\thn{p}
+t^{ab}(p,-q)\,\delta(q_0-2\ep{p} )\,\thp{p}
\Bigr).
\]
Inserting Eq.(\ref{t22}) into the above equations, the imaginary part
of the density-dependent transverse correlation function is obtained
as
\begin{eqnarray}
{\rm Im}\, {\mit\Pi}_{\rm D}( \mitg^2_+, \mitg^2_-\,)
=4\pi\,\delta(q_0)
\int\!\!d^3p\,\frac{M^{\ast2}+p_y^2}{\ep{p}^2}
\Bigl( \thn{p} + \thp{p}-2\thn{p}\thp{p} \Bigr)
+R_{\nnbar}, 
 \label{eq_im_ph}
\end{eqnarray}
where $R_{\nnbar}$ is given by
\begin{eqnarray}
R_{\nnbar}
= -\,4\pi\int\!\!d^3p\,\frac{\vct{p}^2-p_y^2}{\ep{p}^2}
\left(
\delta(q_0+2\ep{p})\,\thn{p}
+\delta(q_0-2\ep{p})\,\thp{p}
\right).  \label{eq_im_pauli}
\end{eqnarray}
From Eqs.(\ref{eq_re_pauli}), (\ref{eq_im_ph}) and (\ref{eq_im_pauli}),
the density-dependent part of the transverse correlation function
is described as
\begin{eqnarray}
 {\mit\Pi}_{\rm D}( \mitg^2_+ , \mitg^2_-\,)
&=&-\,4\int\!\!d^3p\,\frac{M^{\ast2}+p_y^2}{\ep{p}^2}
\left(
\frac{\thn{p}(1-\thp{p})}{q_0+i\varepsilon}
-\frac{\thp{p}(1-\thn{p})}{q_0-i\varepsilon}
\right) \nonumber \\
\noalign{\vskip4pt}
& & 
-\,4\int\!\!d^3p\,\frac{\vct{p}^2-p_y^2}{\ep{p}^2}
\left(
\frac{\thp{p}}{2\ep{p}-q_0-i\varepsilon}
+\frac{\thn{p}}{2\ep{p}+q_0-i\varepsilon}
\right). \label{eq_corr_ph_pauli}
\end{eqnarray}

The density-independent part of the transverse correlation function is
calculated in the same way. From Eqs.(\ref{mff}) and (\ref{t22}), we
obtain
\begin{eqnarray}
{\mit\Pi}_{\rm F}( \mitg^2_+ , \mitg^2_-\,)  
=4\int\!d^3p\,\frac{\vct{p}^2-p_y^2}{E_p^2}
\left(
\frac{1}{2E_p-q_0-i\varepsilon} 
+\frac{1}{2E_p+q_0-i\varepsilon}
\right). \label{eq_n_nbar}
\end{eqnarray}

The sum of Eqs.(\ref{eq_corr_ph_pauli}) and (\ref{eq_n_nbar})
provides us with the full transverse correlation function
${\mit\Pi}( \mitg^2_- , \mitg^2_+\,)$.
It is also expressed as a sum of contributions from
particle--hole and  N--$\overline{\rm N}$ excitations:
\begin{eqnarray}
 {\mit\Pi}( \mitg^2_+ , \mitg^2_-\,)
&=&
{\mit\Pi}_{\rm ph}( \mitg^2_+ , \mitg^2_-\,)
+
{\mit\Pi}_{\nnbar}( \mitg^2_+ , \mitg^2_-\,),\label{full}
\end{eqnarray}
where each term is described as 
\begin{eqnarray}
{\mit\Pi}_{\rm ph}( \mitg^2_+ , \mitg^2_-\,)
&=&-\,4\int\!\!d^3p\,\frac{M^{\ast2}+p_y^2}{\ep{p}^2}
\left(
\frac{\thn{p}(1-\thp{p})}{q_0+i\varepsilon}
-\frac{\thp{p}(1-\thn{p})}{q_0-i\varepsilon}
\right)\,, \label{eq_corr_ph} \\
\noalign{\vskip4pt}
{\mit\Pi}_{\nnbar}( \mitg^2_+ , \mitg^2_-\,)
&=& 
4\int\!\!d^3p\,\frac{\vct{p}^2-p_y^2}{\ep{p}^2}
\left(
\frac{1-\thp{p}}{2\ep{p}-q_0-i\varepsilon}
+\frac{1-\thn{p}}{2\ep{p}+q_0-i\varepsilon}
\right). \label{eq_corr_nnbar}
\end{eqnarray}

\section{The Excitation Energy of the GT State} \label{sec_ex}
The eigenvalues of the excitation energies are given by the real part
of the dimesic function, 
\begin{eqnarray}
 {\rm det\ Re}\ U=0.
\end{eqnarray}
The excitation energy of the GT state is estimated with use of
the transverse part
of the dimesic function in Eq.(\ref{t}). 
The real part of the transverse correlation
function, which we need in the dimesic
function, is obtained from
Eqs.(\ref{full}) to (\ref{eq_corr_nnbar}).
\subsection{The Excitation Energy in the Nucleon Space}
In this subsection, we will calculate the excitation energy of the GT
state, neglecting perfectly the antinucleon degrees of freedom.
In this case, according to Eqs.(\ref{t}) and (\ref{eq_corr_ph}),
the real part of the transverse dimesic function is described as
\begin{eqnarray}
& &{\rm Re}\,D_{\rm T}=1+\chi_5\,{\rm Re}\,{\mit\Pi}_{\rm ph}
(\mitg^{2}_+\,, \mitg^{2}_-\,)\,,\\
\noalign{\vskip4pt} 
& &{\rm Re}\,{\mit\Pi}_{\rm ph}(\mitg^{2}_+\,, \mitg^{2}_-\,)
=-\,\frac{16\pi}{3}
\frac{Q(\kfn)-Q(\kfp)}{q_0},\label{d}
\end{eqnarray}
where $Q(\kfi)$ is given by,
\begin{eqnarray}
Q(\kfi)&=&\frac{3}{4\pi}\int_0^{\kfi}\!d^3p\,
\frac{M^{\ast2}+p_y^2}{\ep{p}^2} \nonumber \\
&=&\frac{\kfi^3}{3}+2\kfi M^{\ast2}
-2M^{\ast3}\tan^{-1}\frac{\kfi}{M^\ast}. \label{qkf}
\end{eqnarray}
From Re $D_{\rm T}=0$, finally we obtain the relativistic expression
of the excitation energy in nuclei with $\kfn > \kfp$,
\begin{equation}
\omega_0=
\frac{2g_5}{3\pi^2}\Bigl( Q(\kfn)-Q(\kfp) \Bigr).\label{q-q}
\end{equation}

Relativistic effects on Eq.(\ref{q-q}) can be seen more transparently,
by defining Fermi momentum $\kf$ as usual
\begin{eqnarray}
\kfn^3=\frac{2N}{A}\kf^3\,,\qquad \kfp^3=\frac{2Z}{A}\kf^3.\label{k3} 
\end{eqnarray}
These yield a relationship for $(N-Z)/A\ll 1$,
\begin{equation}
\kfn-\kfp\approx \frac{2}{3}\kf\frac{N-Z}{A}.\label{n-p}
\end{equation}
By using the equation:
\begin{equation}
Q'(\kf)= \frac{dQ(\kf)}{d\kf}
 =\frac{\kf^2(3M^{\ast2}+\kf^2)}{M^{\ast2}+\kf^2}
=3\kf^2\left(1-\frac23\vf^2\right),
\qquad \vf=\frac{\kf}{\sqrt{M^{\ast2}+\kf^2}},\label{dqkf}
\end{equation}
we expand $\left(Q(\kfn)-Q(\kfp)\right)$ in Eq.(\ref{q-q})
up to first order of $(\kfn -\kfp)$. Then replacing $(\kfn -\kfp)$
by Eq.(\ref{n-p}), 
we obtain the relativistic expression as
\begin{equation}
\omega_0\approx 
\left(1-\displaystyle{\frac23}\vf^2\right)
 g_5\frac{8\kf^3}{3\pi^2}\frac{N-Z}{2A}.\label{gt_eigen}
\end{equation}
The first factor of the r.h.s., depending on the Fermi velocity $\vf$,
shows relativistic effects on the excitation energy.

In the non-relativistic limit $\vf^2 \ll 1$, Eq.(\ref{gt_eigen})
becomes to be
\begin{eqnarray}
\omega_0 = g_5\frac{8\kf^3}{3\pi^2}\frac{N-Z}{2A}.\label{ngt}
\end{eqnarray}
This result can be also obtained 
without using the approximation Eq.(\ref{n-p}). 
In the non-relativistic limit $\vct{p}^2\ll M^{\ast 2}$,
Eq.(\ref{qkf}) becomes
\begin{eqnarray}
Q(\kfi)\approx \kfi^3.
\end{eqnarray}
This, together with Eqs.(\ref{q-q}) and (\ref{k3}), yields
the same result as Eq.(\ref{ngt}).

Eq.(\ref{ngt}) is just the one obtained previously in 
non-relativistic models with $g_5=g'(f_\pi/m_\pi)^2$\cite{suzuki}.
In relativistic models, the excitation energy of the GT state
in nuclear matter
is thus given by the transverse part of the dimesic function, and
is independent of the pion exchange, even when its energy-dependence
is taken into account.

We will show later that the relativistic factor
$(1-2\vf^2/3)$ in Eq.(\ref{gt_eigen}) stems from
the quenching of the GT strength in the nucleon
space. 
In most of the relativistic models the nucleon effective mass is
about $0.6M$\cite{cc},
which yields $\vf = 0.43$ for $\kf = 1.36{\rm  fm}^{-1}$.
This value implies that we must use a larger  
value of $g'$ by 14\% in the relativistic model
than that in non-relativistic models.  
Non-relativistic models require the value of $g'$
to be about 0.6 in order to reproduce experimental data\cite{ss}.
In this case the relativistic model needs to use $g'=0.68$.

\subsection{Effects of the Pauli Blocking Term}

It is known in the relativistic model that
the antinucleon degrees of freedom play an
important role for some physical quantities\cite{ks2, giai}.
Even in the RPA based on the mean field approximation,
a part of the antinucleon excitations should be taken into account in
order to keep the continuity equation\cite{ks1}.
It is the density-dependent part
in the antinucleon excitations, which is usually called the Pauli
blocking term. Without the Pauli blocking term, for example,
the orbital part of the magnetic moment and
giant multiple resonance states are not described correctly\cite{ks2,
giai}.
In the case of the GT state at $\vct{q}=0$, it is not clear
whether or not the Pauli blocking term should be taken into account.
Let us study, however, their effects on the excitation energy of
the GT state. 

The Pauli blocking term in the present case
is given by the density-dependent parts
of Eq.(\ref{eq_corr_nnbar})
\[
 {\mit\Pi}_{\rm Pauli}( \mitg^2_+ , \mitg^2_-\,)
= 
-\,4\int\!\!d^3p\,\frac{\vct{p}^2-p_y^2}{\ep{p}^2}
\left(
\frac{\thp{p}}{2\ep{p}-q_0-i\varepsilon}
+\frac{\thn{p}}{2\ep{p}+q_0-i\varepsilon}
\right).
\]
Its real part is written as    
\begin{equation}
 {\rm Re}\,{\mit\Pi}_{\rm Pauli}( \mitg^2_+ ,  \mitg^2_-\,)
 = \frac{16\pi}{3}\,\kappa\,,
 \qquad
 \kappa=
 -\,P_{\overline{\rm N}}(\kfn,q_0)-P_{\overline{\rm N}}(\kfp,-q_0),
 \label{realnn}
\end{equation}
where we have defined
\begin{eqnarray}
P_{\overline{\rm N}}(\kf,q_0)=
\frac{3}{4\pi}\int_0^{\kf}\!\frac{d^3p}{\ep{p}^2}
\frac{\vct{p}^2-p_y^2}{2\ep{p}+q_0} .
\end{eqnarray}
For $q_0\ll M^\ast$, as in the GT state, $P_{\overline{\rm N}}$ is
 approximately given by 
\begin{eqnarray}
P_{\overline{\rm N}}(\kf,q_0)
\approx P(\kf)&=&
\frac{3}{4\pi}\int_0^{\kf}\!\frac{d^3p}{\ep{p}^2}
\frac{\vct{p}^2-p_y^2}{2\ep{p}} \nonumber \\
&=&
\ekf^2
\left(
\frac32\vf-\vf^3-\frac34\left(1-\vf^2\right)\log\frac{1+\vf}{1-\vf}
\right) \nonumber \\
&=&\kf^2\,\frac{\vf^3}{5}\left( 1+\frac37\vf^2+\cdots\right).
\label{eq_p_nnbar_approx}
\end{eqnarray}
where $\ekf$ denotes $\sqrt{M^{\ast 2}+\kf^2}$.
In taking the above contribution to Eq.(\ref{d}), the excitation
energy of the GT state is obtained as
\begin{equation}
\omega_0\approx 
\frac{1-\displaystyle{\frac23}\vf^2}
 {1+\displaystyle{\frac{2g_5}{3\pi^2}}\kappa}
g_5\frac{8\kf^3}{3\pi^2}\frac{N-Z}{2A}. \label{eq_gt_eigen_p}
\end{equation}
The result shows that
when we use the values of parameters as mentioned at the end of the
previous subsection, the effect of
the Pauli blocking terms is negligible. Even if we should
take into account the Pauli blocking terms, their effects are less
than 0.5\% on the excitation energy.

\section{The GT Strength} \label{sec_gt}

In this section, first we will discuss the total GT strength in
nuclear matter where we can obtain its analytic expression and
understand the structure of the relativistic model.
Next we will investigate effects of finiteness in nuclei.  

\subsection{The GT Strength in Nuclear Matter}

The total GT strength is calculated by integrating the response
function $R$ over the excitation energy. 
The relationship of the response function to the
correlation function ${\mit\Pi}$ is given by\cite{ks1}
\[
 R =\frac{3}{16\pi^2}\frac{A}{\kf^3}{\rm Im}\, {\mit\Pi}.
\]

First we investigate the total GT strength in the mean field
approximation. For this purpose we can employ the imaginary parts of
Eqs.(\ref{full})  to (\ref{eq_corr_nnbar}).
The total strength for the $\beta^-$ transitions in the nucleon space  
is given by the first term in the parentheses of
Eq.(\ref{eq_corr_ph}), 
\begin{equation}
S_{{\rm ph}}^-=
\frac{3}{4\pi}\frac{A}{\kf^3}
\int\!\!d^3p\,\frac{M^{\ast2}+p_y^2}{\ep{p}^2}
\left(\thn{p}-\thp{p}\right)
=\frac{A}{\kf^3}\Bigl( Q(\kfn)-Q(\kfp)\Bigr).\label{sumph}
\end{equation}
When we expand $Q$ in terms of $(\kfn - \kfp)$ as before,
we obtain the value of the total strength in the nucleon space,
\begin{equation}
S_{\rm ph}^-
 \approx \left( 1-\frac23 \vf^2 \right)2\left(N-Z\right).\label{r-sr}
\end{equation}
In the present definition, Ikeda-Fujii-Fujita sum rule in
non-relativistic models\cite{iff} is written as
\begin{eqnarray}
\mtrix{}{Q_+ Q_-}{} - \mtrix{}{Q_- Q_+}{}
=2(N-Z)\,,\label{sr}
\end{eqnarray}
for
\[
 Q_\pm=\sum_i^A\left(\tau_{\pm}\sigma_{y}\right)_i.
\]
This is nothing but the 
result of the commutation relation:
\[
 \left[\,\tau_+\sigma_y\,,\,\tau_-\sigma_y\,\right] = 2\tau_z.
\]
If we assume
that there is no ground-state correlation,
\begin{eqnarray}
Q_+ \ket{\ } = 0\,,\label{s+}
\end{eqnarray}
we have simply from Eq.(\ref{sr})
\begin{eqnarray}
\mtrix{}{Q_+ Q_-}{} = 2(N-Z)\label{non-sr}
\end{eqnarray}
in non-relativistic models.
Comparing Eq.(\ref{r-sr}) with the above equation, it is seen that
the relativistic sum value is quenched by the factor
$( 1-2\vf^2/3 )$, which is about 0.88 for the previous value $\vf
=0.43$.

The strength of the $\beta^+$ transition in the nucleon space
is given by the
second term in the parentheses of Eq.(\ref{eq_corr_ph}) with replacing
$q_0$ by $-q_0$, but its value is zero for $\kfn > \kfp$ as in
Eq.(\ref{s+}).
The quenched strength in the nucleon space in Eq.(\ref{r-sr})
is not taken by the $\beta^+$ transition, but
by the antinucleon degrees of freedom. This fact is shown as follows.
According to the first term of Eq.(\ref{eq_corr_nnbar}),
the strength of the $\beta^-$ transition in the nucleon-antinucleon
excitations is given by
\begin{eqnarray}
S_{\nnbar}^-=\frac{3}{4\pi}\frac{A}{\kf^3}
\int\!\!d^3p\,\frac{\vct{p}^2-p_y^2}{\ep{p}^2}
\left(1-\thp{p}\right),
\end{eqnarray}
while the one of the $\beta^+$ transitions is provided by the
second term with replacing $q_0$ by $-q_0$,
\begin{eqnarray}
S_{\nnbar}^+=\frac{3}{4\pi}\frac{A}{\kf^3}
\int\!\!d^3p\,\frac{\vct{p}^2-p_y^2}{\ep{p}^2}
\left(1-\thn{p}\right).
\end{eqnarray}
The above two equations are both divergent, but their difference is
finite as
\begin{eqnarray}
S_{\nnbar}^- - S_{\nnbar}^+
 = \frac{3}{4\pi}\frac{A}{\kf^3}
\int\!\!d^3p\,\frac{\vct{p}^2-p_y^2}{\ep{p}^2}
\left(\thn{p}-\thp{p}\right).\label{sumnn}
\end{eqnarray}
The sum of the above equation and Eq.(\ref{sumph}) provides us with sum
rule corresponding to Eq.(\ref{sr}),
\begin{eqnarray}
S_{{\rm ph}}^- + S_{\nnbar}^- - S_{\nnbar}^+
 = 2(N-Z).
\end{eqnarray}

In order to obtain the sum rule value $2(N-Z)$, we need a complete set
of the nuclear wave functions. This fact requires both the nucleon
and the antinucleon space in relativistic models. Since the
nucleon-antinucleon states are in the time-like region, the GT
strength for charge-exchange reactions which excite nuclear states
in the space-like region is quenched by the amount of
Eq.(\ref{sumnn}). This quenching can be also discussed by
calculating GT matrix elements directly, as we have
done in ref.\cite{ksg}.

Next we calculate the strength of the GT state in RPA, using
the RPA correlation function
${\mit\Pi}_{\rm RPA}( \mitg^2_+ ,\mitg^2_-\,)$.
When using the abbreviations
${\mit\Pi}_{\rm RPA}( \mitg^2_+,\mitg^2_-\,)
={\mit\Pi}_{\rm RPA}( q_0 )$ and
${\mit\Pi}( \mitg^2_+ ,\mitg^2_-\,)={\mit\Pi}(q_0)$,
${\mit\Pi}_{\rm RPA}( q_0 )$
 is written as
\[
 {\mit\Pi}_{\rm RPA}(q_0)
 =\frac{{\mit\Pi}(q_0)}{D_{\rm T}(q_0)}\,, \qquad
 D_{\rm T}(q_0)=1+\chi_5{\mit\Pi}(q_0).
\]
Expanding $D_{\rm T}(q_0)$ at $q_0=\omega_0$, we have
\[
{\mit\Pi}_{\rm RPA}(q_0)
=
\left(\frac{dD_{\rm T}}{d\omega_0}\right)^{-1}
\frac{{\mit\Pi}(q_0)}{q_0-\omega_0+i\varepsilon}.
\]
When keeping only the density-dependent part of the correlation
function, the imaginary part of the above equation gives the
strength of the GT state,
\begin{equation}
 S_{\rm GT}=\frac{A}{\kf^3}
 \frac{Q(\kfn)-Q(\kfp)}{\Bigl(1+2g_5\kappa/(3\pi^2)\Bigr)^2}
 \approx
 \frac{1-2\vf^2/3}{\Bigl(1+2g_5\kappa/(3\pi^2)\Bigr)^2}
 2\left(N-Z\right). \label{eq_gt_st}
\end{equation}
The $\kappa$-dependent term stems from the Pauli blocking
effects, and is negligible, as mentioned before. Thus in the present
model, the GT state
exhausts the total strength in the nucleon space.  
Comparing Eq.(\ref{q-q}) with the above equation, we can see that
the factor $(1-2\vf^2/3)$ in the expression of the excitation energy
Eq.(\ref{gt_eigen}) is due to the quenching of the GT
strength in the nucleon space, but not from the relativistic
kinematics.

\subsection{The GT Strength in Finite Nuclei}

We have shown analytically that the GT strength, which is responsible
for the giant GT resonance, is quenched by about 12\% in nuclear
matter. The quenched amount, however, depends on the momentum
distribution and the value of the nucleon effective mass near the
nuclear surface, as seen in Eq.(\ref{sumph}). Therefore  
let us estimate numerically the GT strength for finite nuclei
in the mean field approximation. 

We write the four-component nucleon spinor as 
\[
 \psi_{am}=\left(
\begin{array}{c}
\displaystyle{\frac{iG_a(r)}{r}}\,
 \ket{\ell j m} \\
 \noalign{\vskip8pt}
 -\,\displaystyle{\frac{F_a(r)}{r}}\,
 \ket{\bar{\ell} j m} \\
\end{array}
 \right)\,,
\]
where $a$ stands for the quantum numbers $\{n\ell j\}$\,, and
$\bar{\ell}$ is given by $\bar{\ell}=j\pm1/2=\ell\pm1$ for
$j=\ell\pm1/2$.
We define the GT strength  as follows,
\begin{eqnarray*}
T_{aa'}(\sigma_\mu)&=&
2\sum_{mm'} \left|\mtrix{a'm'}{\sigma_\mu}{a\,m}\right|^2
=\frac23\left|\rmtrix{\ell' j'}{\sigma}{\ell j}_{\rm rel}\right|^2,
\end{eqnarray*}
using the notations:
\[
\rmtrix{\ell' j'}{\sigma}{\ell j}_{\rm rel} =
 \delta_{\ell\ell'}
\rmtrix{\ell j'}{\sigma}{\ell j} \,g(a,a')
+
\delta_{\bar{\ell}\bar{\ell}'}
\rmtrix{\bar{\ell}j'}{\sigma}{\bar{\ell}j} \,f(a,a'),
\]
\[
 g(a:a')=\int_0^\infty\!dr\,G_a(r)G_{a'}(r)\,,\qquad
 f(a:a')=\int_0^\infty\!dr\,F_a(r)F_{a'}(r).
\]

If we calculate the strengths for the
transition from $j=\ell+1/2$ to $j'=\ell\pm 1/2$ ( $n'=n$ ) only,
as in non-relativistic models for subshell closed shell nuclei, 
the sum of the GT strengths is given by
\begin{eqnarray*}
\sum_{a' }T_{aa'}(\sigma_\mu)
&=& 
 \frac{4(\ell+1)(2\ell+3)}{3(2\ell+1)}
\left( g_+ - \frac{2\ell+1}{2\ell+3}\,f_+
\right)^2
+ \frac{16\ell(\ell+1)}{3(2\ell+1)}\,g_-^2,
\end{eqnarray*}
with
\begin{eqnarray*}
g_\pm=g(n,\ell,\ell\!+\!1/2:n,\ell,\ell\!\pm\!1/2)\,,\qquad
f_\pm = f(n,\ell,\ell\!+\!1/2:n,\ell,\ell\!\pm\!1/2).
\end{eqnarray*}
In assuming the proton wave function is the same as the neutron wave
function, we have $ g_+ + f_+=1$ from the normalisation of the wave
functions.
Moreover, it may be reasonable to
assume that $g_-\approx 1 - f_+$. Then, the sum of the GT
strengths is approximately given by
\begin{eqnarray*}
\sum_{a' }T_{aa'}(\sigma_\mu)
&\approx&
2(2j+1)
\left( 1-\frac83\,f_+ \right).
\end{eqnarray*}
Since most of the relativistic models provides us with
$f_+\sim 0.02$, the above equation shows that the GT strength is
quenched by about 5\%, compared with the non-relativistic sum 
value $2(2j + 1)$.

In relativistic models, there are other transitions even in sub-shell
closed nuclei, like $^{48}$Ca.
Table \ref{table2} shows that their
contributions to the total GT strength. 
In order to calculate the GT strengths, we have employed the
relativistic model which is named NL-SH\cite{nlsh}.
We calculate the only strength between the bound states.
Contributions from the continuum states are expected to be small.
In the Table, the top one shows the results in using the neutron
wave functions for the initial and final state, and the bottom one
those obtained using the proton wave functions for the final
states. These calculations are performed
to see effects of the Coulomb force. The non-relativistic
sum value for $^{48}$Ca is 16 in the present definition. The Table
shows that the relativistic sum value is quenched by about 6\%,
compared with the non-relativistic one. This reduction of the
quenched amount, compared with the one in nuclear matter, was
expected from the value of the nucleon effective mass near the
nuclear surface, as mentioned before. 
Since the total GT strength in the nucleon space is quenched in the
mean field approximation, we expect that the sum of the RPA strengths
in finite nuclei is also quenched,
as in the case of nuclear matter.

\section{The Pion- and Time-Part of the Correlation Function}

\label{sec_pi}

From section \ref{sec_rel} to the last section we have discussed
the problems related to the transverse part of the correlation
function. In this section let us briefly discuss the structure of
the the pion- and time-component, mainly in order to study the way
to use $g'$ in relativistic models. 

Since at $\vct{q}=0$, the correlation functions satisfy
\begin{eqnarray*}
{\mit\Pi}(\mitg^{-1}_+\,, \mitg^{-1}_-\,)
=q_0^2{\mit\Pi}(\mitg^{0}_+\,, \mitg^{0}_-\,)\,,\qquad
{\mit\Pi}(\mitg^{-1}_+\,, \mitg^{0}_-\,)
=q_0{\mit\Pi}(\mitg^{0}_+\,, \mitg^{0}_-\,),
\end{eqnarray*}
the pion-and time-component of the dimesic function
Eq(\ref{l}) is rewritten in terms of
${\mit\Pi}(\mitg^{0}_+\,, \mitg^{0}_-\,)$,
\begin{eqnarray*}
 D_{\rm PT}
=
1-\frac{1}{(2\pi)^3}\left(
g_5 + g_\pi^2 \frac{q_0^2}{m_\pi^2-q_0^2}
\right){\mit\Pi}(\mitg^{0}_+\,, \mitg^{0}_-\,).
\end{eqnarray*}
The function $t^{00}(p,q)$  at $\vct{q}=0$ in
${\mit\Pi}(\mitg^{0}_+\,,\mitg^{0}_-\,)$ 
is calculated according to Eq.(\ref{t1}) as,
\begin{eqnarray*} 
t^{00}(p,q)=
4\left( 2M^{\ast2}-\ep{p}( 2\ep{p}+q_0 ) \right).
\end{eqnarray*}
In taking into account the density--dependent parts only,
we have the real part of the time--component as  
\begin{eqnarray}
{\rm Re}{\mit\Pi}(\mitg^{0}_+\,,\mitg^{0}_-\,)
 =J^{00}(\kfn,q_0)+J^{00}(\kfp,-q_0),
\end{eqnarray}
where $J^{00}$ is given by
\begin{eqnarray}
J^{00}(\kfi,q_0)
=-\,\frac{4}{q_0}\int\!d^3p\,\frac{\vct{p}^2}{\ep{p}^2}\thi{p}
-4M^{\ast2}\int\!d^3p\,\frac{\thi{p}}{\ep{p}^2(2\ep{p}+q_0)} .
\label{eq_long1}
\end{eqnarray}
For $q_0\ll M^\ast $, neglecting the $q_0$--dependence of the second
term, we obtain
\begin{eqnarray}
J^{00}(\kfi,q_0)\approx 
8\pi M^{\ast2}
\left( \frac{Q_0(\kfi)}{q_0}+P_0(\kfi) \right),\label{j00}
\end{eqnarray}
with
\begin{eqnarray}
P_0(\kf)&=& 
-\,\frac{1}{2\pi}\int_0^{\kf}\!d^3p\,\frac{1}{2\ep{p}^3}
=
\frac{\kf}{\ekf}-\log\frac{\ekf+\kf}{M^\ast}
=-\,\frac{\vf^3}{3}\left(1+\frac{3}{5}\vf^2+\cdots\right),\label{p2}
 \\
\noalign{\vskip4pt}
Q_0(\kf)&=&
-\,\frac{1}{2\pi M^{\ast2}}\int_0^{\kf}\!d^3p\,
\frac{\vct{p}^2}{\ep{p}^2}
=2\kf-2M^\ast\tan^{-1}\frac{\kf}{M^\ast}
-\frac{2\kf^3}{3M^{\ast2}}\,.
\end{eqnarray}
From the above equations, the real part of
 ${\mit\Pi}(\mitg^{0}_+\,,\mitg^{0}_-\,)$ is described as
\begin{eqnarray}
{\rm Re}\ {\mit\Pi}(\mitg^{0}_+\,,\mitg^{0}_-\,)
\approx 16\pi M^{\ast2}\left(P_0(\kf)
+\frac{\kfn-\kfp}{2q_0}Q_0'(\kf)\right)\,,\label{lrp}
\end{eqnarray}
where $Q_0'(\kf)$ denotes the derivative of $Q_0(\kf)$
with respect to $\kf$,
\begin{eqnarray}
Q_0'(\kf)=-\,\frac{2\kf^4}{M^{\ast2}\ekf^2}
=-\,\frac{2\vf^4}{1-\vf^2}.\label{q2}
\end{eqnarray}
Finally the real part of the PT dimesic function is given by
\begin{eqnarray}
{\rm Re}\ D_{\rm PT}
\approx 1-\frac{2M^{\ast2}}{\pi^2}
\left( g_5+g_\pi^2
 \frac{q_0^2}{m_\pi^2-q_0^2}
\right)
 \left(
 P_0(\kf)+\frac{\kfn-\kfp}{2q_0}Q'_0(\kf)
 \right).\label{ld2}
\end{eqnarray}

The structure of ${\rm Re}\ D_{\rm PT}$ is similar
to ${\rm Re}\ D_{\rm T}$ in Eq.(\ref{d}) to which Eq.(\ref{realnn})
is added.
Eqs.(\ref{p2}) and (\ref{q2}), however, show that the quantity in the
second parenthesis in the above equation is negative. Therefore,
the excitation
energy given by ${\rm Re}\ D_{\rm PT}=0$
should be higher than the pion mass $q_0 >m_\pi$. 

The first parenthesis of Eq.(\ref{ld2}) may be obtained by 
the insertion of $g'$ into the pion propagator as
\begin{equation}
 \frac{1}{m_\pi^2-q^2}\longrightarrow
 \frac{1}{m_\pi^2-q^2}+\frac{g'}{q^2},\label{rep1}
\end{equation}
which was frequently used in relativistic description of
 high-momentum transfer reactions\cite{h}.
Eq.(\ref{ld2}), however, shows that the way to put $g'$ in the
meson propagator  cannot describe the GT states.
In order to show this fact, we have used the Lagrangian Eq.(\ref{L}),
although the GT state can be described only by the
contact term in nuclear matter.
The above modification of the meson propagator in Eq.(\ref{rep1})
was introduced from non-relativistic models. Those models 
use a static potential and modify the meson propagator so as to
cancel the short range part of the interaction as
\begin{equation}
 \frac{\vct{q}^2}{m_\pi^2+\vct{q}^2} \longrightarrow
 \frac{\vct{q}^2}{m_\pi^2+\vct{q}^2}-g'.\label{rep2}
\end{equation}
Eq.(\ref{rep1}), however, is not a reasonable
extension of Eq.(\ref{rep2}) for
description of the GT state. 

The last statement, of course, does not mean that the Lagrangian
form in Eq.(\ref{L}) provides us with a correct four-momentum
dependence of $g'$. In non-relativistic models also we do not
know the dependence so well\cite{dik}.
The Lagrangian form in Eq.(\ref{L}) can describe the GT state
at $\vct{q}=0$, and cancel the short range part of
the interaction, but yields an additional four-momentum transfer
dependence of the dimesic function. In fact, the dimesic function
except for the transverse part is written at the static
limit $q_0=0$ as,
\begin{eqnarray}
-D_{\rm PTL}
 = (1-\chi_5{\mit\Pi}^{00})
 \left(1+(\chi_5-\chi_\pi q_x^2){\mit\Pi}^{11}\right)
+ \chi_5(\chi_5-\chi_\pi q_x^2)({\mit\Pi}^{10})^2,
\end{eqnarray}
where we have used the abbreviation:
${\mit\Pi}^{ab}={\mit\Pi}( {\mit\Gamma}^a_+ , {\mit\Gamma}^b_-\,)$.
More detailed investigation on $g'$ is necessary
for discussions of high momentum transfer phenomena.

Finally we note effects of the Pauli blocking terms.
When we take into account the only particle-hole excitations, we have
${\rm Re}\ {\mit\Pi}(\mitg^{0}_+\,,\mitg^{0}_-\,)$ as
\begin{eqnarray*}
{\rm Re}\ {\mit\Pi}(\mitg^{0}_+\,,\mitg^{0}_-\,)
=\frac{1}{q_0+i\varepsilon}
\int\!\!d^3p\,\frac{t^{00}}{2\ep{p}^2}
\Bigl( \thn{p} - \thp{p} \Bigr)
=\frac{8\pi M^{\ast2}}{q_0+i\varepsilon}\,
\Bigl( Q_0(\kfn) - Q_0(\kfp) \Bigr)
\end{eqnarray*}
This shows that the term $P_0(\kf)$ in Eq.(\ref{lrp}) comes from
the Pauli blocking terms.
The effects of the Pauli blocking terms are
not small in the present case,
compared with those in the transverse mode. In fact,
the relationship between contributions from the particle-hole
term to the Pauli blocking one is given by
\[
 Q_0'(\kf)\approx  -\,2\vf^4\,,\qquad
 P_0(\kf)\approx -\,\frac{\vf^3}{3}
 \approx \frac{1}{6\vf}\,Q_0'(\kf), 
\]
in the present case,
while in the transverse mode, we have from Eqs.(\ref{dqkf})
and (\ref{eq_p_nnbar_approx})
\[
 Q'(\kf)\approx 3\kf^2\,,\qquad
 P(\kf)\approx \kf^2\,\frac{\vf^3}{5}\approx \frac{\vf^3}{15}\,Q'(\kf).
\]

\section{Summary}

\label{sec_sm}

In 1980's, analytic expressions of the excitation energies for
the  giant monopole and quadrupole resonance states were 
derived in the relativistic model\cite{nks}.
When they are expressed in terms of
the Landau-Migdal(LM) parameters, they are formally equal to the
non-relativistic expressions, in spite of the fact that
the LM parameters are strongly dominated by relativistic effects.
In this paper, we have obtained the
relativistic expression of the excitation energy for the 
Gamow-Teller(GT) state in nuclear matter.
It is described in terms of the LM
parameter $g'$ which is introduced in the Lagrangian as a contact
term. Compared with the corresponding non-relativistic one,
the relativistic expression has an additional factor of
$(1-2\vf^2/3)$, $\vf$ being the Fermi velocity.
This means that in order to reproduce the same excitation energy as
in non-relativistic models, the present relativistic model requires a
larger value of $g'$  by this factor.

The above relativistic factor comes from the quenching of the GT
strength in the nucleon space. A part of the GT strength is taken by
the nucleon-antinucleon states in the time-like region which
are not excited in usual charge-exchange reactions.
This quenching is thus peculiar to the relativistic models.  
The quenched amount is estimated to be 12\% of the
classical Ikeda-Fujii-Fujita sum rule value in nuclear matter,
and 6\% in finite nuclei.

Recently experiment has
observed 90\% of the classical sum rule value in
$^{90}$Zr\cite{sakai}, although the data were analysed in
non-relativistic models. So far the quenching of 10\% has been
considered to
be due to the coupling of the particle-hole states with $\Delta$-hole
states. Under this assumption, the LM parameter $g'_{\Delta {\rm N}}$
for the coupling is estimated to be about 0.2 to 0.3, depending on
the model\cite{ss}.
The determination of the value of $g'_{\Delta {\rm N}}$
is very important for studies of nuclear magnetic moments
and pion condensation. In particular,
the critical density of the pion condensation is dominated by the
value of $g'_{\Delta {\rm N}}$. It has been shown that if its value
is about 0.2, a rough calculation yields the critical density to be
about 2 times of the normal density\cite{sst}.    
In the present
relativistic model the
nucleon-antinucleon excitations are also responsible for the 
quenching. If a half of the quenching is owing to the
nucleon-antinucleon excitations, the value of $g'_{\Delta {\rm N}}$
becomes to be about a half of the above value, and consequently the
critical density becomes lower. More detailed investigation on the
observed quenching is required in the relativistic model.

In this paper we have also discussed whether or not
it is appropriate for the relativistic model
to insert the LM parameter $g'$
into the meson propagator.
This method was frequently employed for the study of
high-momentum reactions, but we have shown that this method can not
describe the GT states.

Furthermore, it has been 
shown that the Pauli blocking terms are not important
for discussions of the GT states.

\begin{acknowledgments}
The authors would like to thank Professor Z. Y. Ma for
useful discussions.
\end{acknowledgments}

\newpage

\begin{table} 
\caption{\label{table1}The structure of the correlation function
 ${\mit\Pi}( \mitg^a_- , \mitg^b_+\,)$.
The first column and row indicate the values of $a$
and $b$ of ${\mit\Pi}( \mitg^a_- , \mitg^b_+\,)$.
The open boxes mean that ${\mit\Pi}( \mitg^a_- , \mitg^b_+\,)$
has non-zero value. Table (a) is for $\vct{q}\neq 0$, while
(b) for $\vct{q}= 0$.}
\begin{tabular}{|r||c|c|c|c|c|}
\multicolumn{6}{c}{(a)}\\[4pt]
 \hline
     & $-1$& \ 0\ \  & \ 1\ \  & \ 2\ \  & \ 3\ \ \\ \hline\hline
$-1$ &     &   &   & 0 & 0 \\ \hline
 0   &     &   &   & 0 & 0 \\ \hline
 1   &     &   &   & 0 & 0 \\ \hline
 2   & 0   & 0 & 0 &   & 0 \\ \hline
 3   & 0   & 0 & 0 & 0 &  \\ \hline
 \end{tabular}
\qquad
\begin{tabular}{|r||c|c|c|c|c|}
\multicolumn{6}{c}{(b)}\\[4pt]  \hline
     & $-1$& \ 0\ \  & \ 1\ \ &\ 2\ \ & \ 3\ \ \\ \hline\hline
$-1$ &     &   & 0 & 0 & 0 \\ \hline
 0   &     &   & 0 & 0 & 0 \\ \hline
 1   & 0   & 0 &   & 0 & 0 \\ \hline
 2   & 0   & 0 & 0 &   & 0 \\ \hline
 3   & 0   & 0 & 0 & 0 &  \\ \hline
 \end{tabular}  
 \end{table}

\begin{table}
\caption{
The GT strength of the single-particle transition in $^{48}$Ca.
The top table shows the results
obtained using neutron wave functions for the initial and final
states, while the bottom one those using the proton wave functions
for final states.
The value in the parentheses following the single-particle quantum
number shows the binding energy in MeV. $T_{aa'}$ is the value of
the GT strength, and 
$g(a:a')$ and $f(a:a')$ show the overlap of the radial wave functions,
 as defined in the text. The values indicated by the underline do not
 contributed to the GT strength.}
\label{table2}
\begin{tabular}{|ll|r|r|r|}
\multicolumn{5}{c}{\large n\,$\rightarrow$\,n} \\\hline
$n\ell j$ & $n'\ell'j'$ & $T_{aa'}$ & $g(a:a')$ & $f(a:a')$\\\hline
$1p_{3/2}(-39.41)$ & $2p_{3/2}( -3.97)$& $ 0.001$ & $ 0.0110$ &
 $-0.0110$ \\
$1p_{3/2}(-39.41)$ & $1f_{5/2}( -2.09)$ & $ 0.002$ &
 \underline{$ 0.7765$} & $-0.0175$ \\
$1p_{3/2}(-39.41)$ & $2p_{1/2}( -2.74)$ & $ 0.004$ & $ 0.0323$ &
 \underline{$-0.0091$}\\
$1p_{1/2}(-36.23)$ & $2p_{3/2}( -3.97)$ & $ 0.001$ & $-0.0128$ &
 \underline{$ 0.0116$}\\
$1p_{1/2}(-36.23)$ & $2p_{1/2}( -2.74)$& $ 0.001$ & $ 0.0102$ &
 $-0.0102$ \\
$1f_{7/2}(-10.00)$ & $1f_{5/2}( -2.09)$ & $ 8.411$ & $ 0.9592$ &
 \underline{$-0.0150$}\\
$1f_{7/2}(-10.00)$ & $1f_{7/2}(-10.00)$& $ 6.390$ & $ 0.9805$ &
 $ 0.0195$ \\
\hline
 &Total & $14.810$ & & \\
\hline\end{tabular}

\vspace{\baselineskip}
 
\begin{tabular}{|ll|r|r|r|}
\multicolumn{5}{c}{\large n\,$\rightarrow$\,p} \\\hline
$n\ell j$ & $n'\ell'j'$ & $T_{aa'}$ & $g(a:a')$ & $f(a:a')$\\\hline
$1p_{3/2}(-39.41)$ & $2p_{3/2}( -1.09)$& $ 0.000$ & $-0.0113$ &
 $-0.0116$ \\
$1p_{3/2}(-39.41)$ & $1f_{5/2}( -1.16)$ & $ 0.002$ &
\underline{$ 0.8084$} & $-0.0180$ \\
$1p_{1/2}(-36.23)$ & $2p_{3/2}( -1.09)$ & $ 0.005$ & $-0.0360$ &
 \underline{$ 0.0117$}\\
$1f_{7/2}(-10.00)$ & $1f_{5/2}( -1.16)$ & $ 8.629$ & $ 0.9715$ &
 \underline{$-0.0148$}\\
$1f_{7/2}(-10.00)$ & $1f_{7/2}( -9.59)$& $ 6.361$ & $ 0.9787$ &
 $ 0.0200$ \\ 
\hline
 &Total & $14.997$ & & \\
\hline
\end{tabular}
\end{table}


\begin{thebibliography}{99}
\bibitem{swring}B. D. Serot and J. D. Walecka,
	Adv. Nucl. Phys. {\bf 16},
	ed. E. Vogt and J. Negle (Plenum, N. Y., 1986); P. Ring,
	Prog. Part. Nucl. Phys. {\bf 37} (1996) 193.
\bibitem{conti}C. De Conti, A. P. Gale$\tilde{{\rm a}}$o and F.
	Krmpoti$\acute{{\rm c}}$,
	Phys. Lett. {\bf B444} (1998) 14; {\bf B494} (2000) 46.
\bibitem{h}
	J. F. Dawson and J. Piekarewicz, Phys. Rev. {\bf C43} (1991)
	2631; C. J. Horowitz and J. Piekarewicz, Phys. Rev. {\bf C50}
	(1994) 2540.
\bibitem{mosel}M. Sch$\ddot{\rm a}$fer, H. C. D$\ddot{\rm o}$nges,
	A. Engel and U. Mosel,
	Nucl. Phys. {\bf A575} (1994) 429.
\bibitem{ks1}
	H. Kurasawa and T. Suzuki, Nucl. Phys. {\bf A445} (1985) 685.
\bibitem{chin}S. A. Chin, Ann. of Phys. {\bf 108} (1977) 301.
\bibitem{suzuki}T. Suzuki, Nucl. Phys. {\bf A379} (1982) 110.
\bibitem{cc}M. Centelles, M. Del Estal and X. Vi$\tilde{{\rm n}}$as,
	Nucl. Phys. {\bf A635} (1998) 193.
\bibitem{ss}T. Suzuki and H. Sakai, Phys. Lett. {\bf B455} (1999) 25;
	A. Arima, W. Bentz, T. Suzuki and T. Suzuki, Phys. Lett. {\bf
	B499} (2001) 104, and references therein.
\bibitem{ks2}H. Kurasawa and T. Suzuki, Phys. Lett. {\bf B474} (2000) 262.
\bibitem{giai}Z. Ma, N. V. Giai, A. Wandelt, D. Vretenar and P. Ring,
	Nucl. Phys. {\bf A686} (2001) 173. 
\bibitem{iff}K. Ikeda, S. Fujii and J. I. Fujita, Phys. Lett. {\bf 3}
	(1963) 271.
\bibitem{ksg}H. Kurasawa, T. Suzuki and N. V. Giai, nucl-th/0301074.
\bibitem{nlsh} M. M. Sharma, M. A. Nagarajan and P. Ring,
	Phys. Lett. {\bf B312} (1993) 377.
\bibitem{dik}W. H. Dickhoff et al., Phys. Rev. {\bf C23} (1981) 1154.
 \bibitem{nks}S. Nishizaki, H. Kurasawa and T. Suzuki, Nucl. Phys. {\bf
	A462} (1987) 687.
\bibitem{sakai}T. Wakasa et al., Phys. Rev. {\bf C55} (1997) 2909.
\bibitem{sst}T. Suzuki, H. Sakai and T. Tatsumi, in Proc. of the RCNP
	Int. Symp. on Nuclear Responses and Medium Effects (~Universal
	Academic Press,Tokyo, 1999 ) 77.
\end{thebibliography}
\end{document}